\documentclass[aps,prd,nofootinbib,twocolumn,superscriptaddress,letterpaper,preprintnumbers,notitlepage]{revtex4-1}

\usepackage{amsmath,amssymb,braket,mathtools}
\usepackage[dvipsnames]{xcolor}
\usepackage{color}
\usepackage{float}
\usepackage{graphicx}
\usepackage{subfigure}
\usepackage{natbib}
\usepackage{adjustbox}
\usepackage[colorlinks=true
,urlcolor=blue
,anchorcolor=blue
,citecolor=blue
,filecolor=blue
,linkcolor=red
,menucolor=blue
,hyperfootnotes=false,
,linktocpage=true
,pdfproducer=medialab
,pdfa=true
]{hyperref}

\newcommand{\n}{\nonumber \\}

\newcommand\orcid[1]{\href{http://orcid.org/#1}{\adjustbox{trim={-.15\width} {0\height} {-.15\width} {0\height},clip}{\includegraphics[height=10pt]{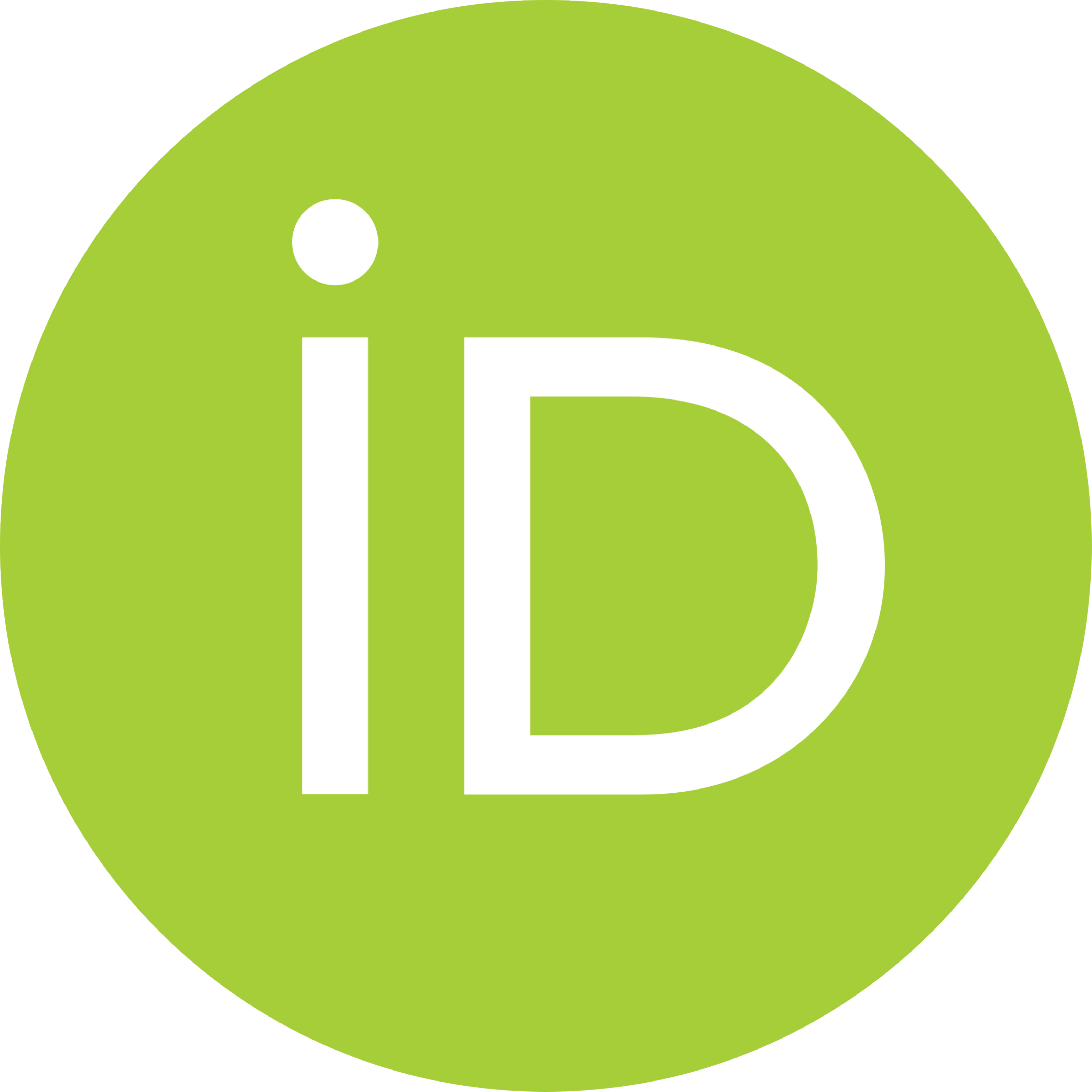}}}}

\begin{document}

\title{Enhanced density fluctuations: \\
consequences for stellar dynamics and dark matter substructure}

\author{Wenzer Qin\orcid{0000-0001-7849-6585}}
\affiliation{Center for Cosmology and Particle Physics, Department of Physics, New York University, New York, NY 10003, USA}

\author{David E. Kaplan\orcid{0000-0001-8175-4506}}
\affiliation{William H. Miller III Department of Physics and Astronomy, Johns Hopkins University, 3400 N. Charles Street, Baltimore, Maryland, 21218, USA}

\author{Neal Weiner\orcid{0000-0003-2122-6511}}
\affiliation{Center for Cosmology and Particle Physics, Department of Physics, New York University, New York, NY 10003, USA}

\begin{abstract}
    Enhanced density fluctuations form early dark matter halos, which serve as early gravitational potential wells for baryons.
    These baryons can then condense to form stars and black holes at high redshifts, $1+z \gtrsim 100$.
    Larger density fluctuations result in denser halos with higher virial temperatures; hence, a halo at a mass that is below the threshold for forming stars under standard astrophysical expectations may form astrophysical objects if its initial density is higher than expected, although the exact threshold for compact object formation is subject to large astrophysical uncertainties.
    We outline the conditions under which enhanced density fluctuations form dark (star-less) halos, early galaxies, or even star clusters.
    Using toy model enhancements to the primordial power spectrum that have been shown to explain the abundances of ``little red dots", along with our fiducial set of astrophysical parameters, we show that the number of clusters initially captured by the Milky Way is highly sensitive to the shape of the enhancement, with the narrowly peaked enhancement resulting in $50$ clusters and the step-function enhancement resulting in $\sim 4.1 \times 10^5$.
    The initial stellar masses of the clusters is also $\mathcal{O} (100) \,M_\odot$, although this range is also dependent on the shape of the enhancement.
    In addition, the abundance of dark matter subhalos with masses greater than $10^7 \, M_\odot$ can be enhanced by as much as a factor of $\sim 6.5$ and is denser than in $\Lambda$CDM, which will increase their detectability in future stellar stream surveys such as the Via Project.
\end{abstract}

\maketitle

\section{Introduction}

The small scale matter power spectrum is a potentially powerful probe of cosmology and the fundamental nature of dark matter, yet it remains relatively unconstrained for scales smaller than about ten megaparsecs (or halos smaller than $10^9 \, M_\odot$)~\cite{Bechtol:2022koa,Sabti:2021unj}.
Within the standard $\Lambda$CDM lore, halos much smaller than this are either incapable of forming stars because their virial temperature is too small, or if they do form stars, then supernova feedback may eject the luminous matter and leave the halos still dark~\cite{2011ARA&A..49..373B}.
However, if the matter power spectrum increases towards small scales which have not yet been observed, then this can lead to the formation of earlier and denser dark matter halos which have high virial temperatures and are robust to stellar feedback, even at relatively small halo masses.
The higher density of the dark matter can also lead to richer stellar dynamics.

Many exotic particle physics models predict enhancements in power, including, but not limited to, models of inflation with phases of ``ultra slow-roll"~\cite{Kinney:2005vj,Ezquiaga:2017fvi,Garcia-Bellido:2017mdw,Kannike:2017bxn,Motohashi:2017kbs,Germani:2017bcs,Di:2017ndc,Ballesteros:2017fsr,Mishra:2019pzq,Karam:2022nym,Ozsoy:2023ryl,Cole:2023wyx,Cicoli:2018asa,Bhaumik:2019tvl,Choudhury:2024one,Cai:2022erk,Inomata:2021tpx,Geller:2022nkr}, cosmologies with periods of early matter domination~\cite{Erickcek:2011us,Erickcek:2015jza,Barenboim:2013gya,Blanco:2019eij,Inomata:2019ivs,Co:2025lrd}, and axion misalignment~\cite{Hogan:1988mp,Fairbairn:2017sil,Buschmann:2019icd,Blinov:2019jqc,Arvanitaki:2019rax}.
The larger variance in density causes some overdensities to evolve into the nonlinear regime early, which leads to the formation of halos that are denser than within na\"{i}ve $\Lambda$CDM expectations.
This has led to the exploration of, {\it e.g.}, ultra-compact dark matter halos~\cite{Delos:2017thv,Delos:2018ueo}, but the baryonic consequences of these early dense halos are relatively underexplored.

In particular, due to their higher density, these early dark matter halos can be above the threshold for atomic and molecular cooling, even if they are relatively small in mass, and therefore host the formation of astrophysical objects.
Motivated by the observations of ``little red dots" (LRDs) from the \textit{James Webb Space Telescope} (JWST), the authors of Ref.~\cite{Qin:2025ymc} showed that under the right conditions, early-forming halos can seed supermassive black holes (SMBHs) through direct collapse---these early seeds were called ``not-quite-primordial" black holes.
For halos in place prior to redshifts of $z \gtrsim \mathcal{O} (100)$~\cite{2015ApJ...814...18H,2024PASJ...76..850I}, the cosmic microwave background is hot enough to prevent the formation of molecular hydrogen, which is a key condition for direct collapse.
So long as the angular momentum and relative velocity between baryons and dark matter is also small within these halos, then this can lead to the formation of an intermediate mass black hole, which can grow into a supermassive black hole at late redshifts without invoking super-Eddington accretion.

In this work, we aim to reach a more general understanding of the consequences of baryons falling into these early halos.
In particular, if the halos fail to meet the conditions for direct collapse, then the objects forming are not quite not-quite-primordial black holes, and can instead become systems comprised of stars.
To understand the range of observable consequences that this may lead to, we must understand how these systems will be classified if they are observed today.
We identify a few conditions that allow us to categorize the types of astrophysical systems that can form from enhanced overdensities and show that the minimum halo mass at which an overdensity can form and retain stars can be lower than within typical $\Lambda$CDM expectations.
In addition, we show that the high density of the early halos can lead to interesting stellar dynamics and the formation of star clusters, as well as enhanced dark matter substructure.

We proceed in this manuscript as follows: in Section~\ref{sec:early_structure}, we discuss how enhanced density fluctuations evolve and how the high virial temperatures can lead to early stars or SMBH seeds.
In Section~\ref{sec:stellar}, we outline conditions for determining the impact of gas temperature, supernovae feedback, and dynamical friction on the formation of stellar systems, and discuss astrophysical uncertainties.
In Section~\ref{sec:results}, we summarize the results and delineate the regions of parameter space that can result in dark (i.e. star-less) halos, early galaxies, or early star clusters.
In Section~\ref{sec:predictions}, we discuss further predictions for these stellar systems and their associated halos.
Finally, we conclude with Section~\ref{sec:conclude}.

\section{Early structure formation}
\label{sec:early_structure}

Enhanced density fluctuations can collapse earlier than expected in standard cosmology, and form early black holes and stars.
The fate of these large density fluctuations depends primarily on their characteristic wavelength, which determines the resulting halo mass; the initial amplitude, which determines when the overdensity collapses; and also on properties such as the halo's angular momentum and dark matter-baryon relative velocity.
For example, for initial density fluctuations with $\delta \equiv \frac{\rho}{\bar{\rho}} - 1\gtrsim 1/3$, where $\rho$ represents density and $\bar{\rho}$ is the mean density, the overdensities immediately collapse into primordial black holes (PBHs) upon entering the horizon~\cite{Carr:1975qj}.

Density fluctuations 
\footnote{Here, we are assuming adiabatic fluctuations, although dark matter isocurvature fluctuations can also result in early dark matter halos.}
below this threshold instead result in dense dark matter halos that act as early gravitational potential wells.
The virial density of a halo at the redshifts of interest is given by
\begin{equation}
    \rho_\mathrm{vir} = 18 \pi^2 \Omega_m \bar{\rho}_{c,0} (1+z_\mathrm{NL})^3 ,
    \label{eqn:rho_vir}
\end{equation}
where $\Omega_m$ is the matter density parameter, $\bar{\rho}_{c,0}$ is the critical density today and $z_\mathrm{NL}$ is the redshift at which the overdensity exceeds the threshold for collapse ($\delta > 1.69$ within linear evolution), which can be calculated numerically using Boltzmann codes such as \texttt{CLASS}~\cite{2011arXiv1104.2932L,2011JCAP...07..034B}.
From this expression, one can see that halos that form earlier have an increased density.

The subsequent temperature for baryons in the halo's potential is given by the virial temperature
~\cite{2010gfe..book.....M}:
\begin{equation}
    T_\mathrm{vir} \approx 9700 \,\mathrm{K} \, 
    \left( \frac{M_\mathrm{halo}}{10^7 M_\odot} \right)^{2/3} 
    \left( \frac{\rho}{4 \times 10^{-24} \,\mathrm{g}\,\mathrm{cm}^{-3}} \right)^{1/3} ,
    \label{eqn:Tvir}
\end{equation}
where $M_\mathrm{halo}$ denotes the halo mass.
Hence, early-forming halos also have higher virial temperatures that allow them to be above the threshold for atomic cooling, which is approximately $10^4$~K~\cite{Barkana:2000fd,Oh:2001ex}.
Since the baryons are able to cool and condense, they can begin to form astrophysical objects at very high redshifts, even within relatively small halos.

If enough molecular hydrogen forms so that molecular cooling is efficient as the baryons collapse, then the gas can cool down to $\mathcal{O} (100)$~K and will fragment into stars~\cite{2015ApJ...814...18H,2024PASJ...76..850I}.
At high redshifts, the cosmic microwave background (CMB) is hot enough to significantly dissociate molecular hydrogen; hence, star formation only begins to occur when $z < z_\mathrm{crit}$, where the redshift threshold for molecular cooling halos to form has been calculated to be $z_\mathrm{crit} \in [200,  400]$~\cite{2015ApJ...814...18H,2024PASJ...76..850I,Qin:2025ymc}.
Due to the relatively high Jeans mass in molecular cooling halos, these stars can have masses of $\sim 100 -1000 \,M_\odot$~\cite{2024PASJ...76..850I}.

If, instead, the baryons are able to fall into these potential wells at sufficiently early times, before significant amounts of molecular hydrogen can form, then cooling is suppressed and the baryons may undergo direct collapse to become not-quite-primordial black holes (NQPBHs), as pointed out in Ref.~\cite{Qin:2025ymc}.
Not all halos that form above $z_\mathrm{crit}$ are able to undergo direct collapse---if the angular momentum of the baryons or their velocity relative to the dark matter is too high, this can delay the collapse of the baryons until $z < z_\mathrm{crit}$, allowing stars to form even if the halos already are in place at high redshifts. 
In what follows, we will investigate the kinds of astrophysical systems that can form once these halos begin to form stars.

\section{Stellar dynamics}
\label{sec:stellar}

Here we will describe the physics necessary to understand the qualitative properties of the stellar systems resulting from early structure formation.
Namely, we will focus on whether stars are able to form at all, whether the gas and stars that form in a halo are able to remain bound under the effect of supernovae, and also whether the stars and dark matter form in a sufficiently dense initial configuration such that gravitational scatters (and therefore kinetic energy transfer) become efficient---we will also discuss the astrophysical uncertainties relevant to each process.

\subsection{Cooling thresholds}

For stars to be able to form within a halo at all, the gas within the halo's potential must be ``hot enough to cool".
In other words, processes such as molecular and atomic hydrogen cooling are usually efficient \textit{above} a temperature threshold; for atomic hydrogen cooling, this is $\sim 10^4$~K~\cite{Barkana:2000fd,Oh:2001ex}, and for molecular hydrogen cooling, this threshold is $\sim 10^2-10^3$~K~\cite{Barkana:2000fd}.
As discussed in the previous section, a lower temperature threshold corresponds to a smaller fragment mass; hence a halo capable of undergoing atomic cooling can form objects such as intermediate-mass black holes, a halo undergoing molecular cooling will form smaller Pop III stars, and halos below both thresholds in the early universe will not form stars at all.

We would like to know for a given dark matter overdensity with wavenumber $k$ and initial amplitude $\delta_{c,0}$ whether in the infalling gas can achieve virial temperatures high enough to trigger cooling.
Using the fact that the halo mass is
\begin{equation}
    M_\mathrm{halo} \approx \frac{4}{3} \pi \rho \left( \frac{\pi}{k} \right)^3 ,
\end{equation}
as well as Eqs.~\eqref{eqn:rho_vir} and \eqref{eqn:Tvir}, we can obtain $T_\mathrm{vir}$ for any halo as a function of $k$ and $\delta_{c,0}$.
Comparing $T_\mathrm{vir}$ to the temperature thresholds described above then allows us to determine whether the gas in a halo is capable of cooling and forming astrophysical objects in the first place.

\subsection{Unbinding by supernovae}
\label{sec:supernovae}

As stars form within early halos, these stars may eventually become supernovae.
For a dark matter halo with a low enough density, i.e. a shallow gravitational potential, it is possible for the injection of energy by supernovae to unbind the remaining stars and gas within the halo.
We would like to understand where this threshold is, as it affects the demarcation between the overdensities that form dark objects vs. the overdensities that have luminous content today.

Given the mass of a halo $M_\mathrm{halo}$ and its redshift of formation $z_\mathrm{NL}$, we take the gravitational binding energy of a halo to be
\begin{equation}
    E_\mathrm{grav} \approx \frac{G M_\mathrm{halo}^2}{r_\mathrm{vir}} , 
    \label{eq:Egrav}
\end{equation}
where the characteristic radius of the virialized halo is given by
\begin{equation}
    r_\mathrm{vir} = \left( \frac{3 M_\mathrm{halo}}{4 \pi \rho_\mathrm{vir}} \right)^{1/3} .
\end{equation}
We note that Eq.~\eqref{eq:Egrav} may have an additional $\mathcal{O}(1)$ constant factor which depends on the density profile of the halo, but given the larger uncertainties on other astrophysical parameters relevant for this section, this expression for $E_\mathrm{grav}$ is sufficient.

The number of stars that form in a halo is approximately given by
\begin{equation}
    N_* = \left( \frac{\Omega_b}{\Omega_m} M_\mathrm{halo} f_* \right) \frac{1}{M_*}
\end{equation}
where $\Omega_b$ is the baryon density parameter, $f_*$ is the fraction of baryonic mass that is converted to stars, and $M_*$ is the typical mass of an individual star.
Here, we assume an initial baryon fraction equal to the cosmic baryon fraction, although the final baryon fraction may be much smaller than this, as has been shown in the case of low-mass systems and ultrafaint dwarfs~\cite{McGaugh:2009mt,McGaugh:2010yr}.

To approximate the energy released by supernovae, we take the energy of a single core-collapse supernova to be $E_\mathrm{CC} = 10^{51}$~ergs~\cite{Weinberg_2006}.
If we use $f_\mathrm{SN}$ to denote the fraction of stars that go supernova, then the total energy released by supernova is given by
\begin{equation}
    E_\mathrm{SN} = N_* \, f_\mathrm{SN} \, E_\mathrm{CC}.
    \label{eq:ESN}
\end{equation}

Each of these astrophysical factors merits additional discussion. 
First, we take the value of the star formation efficiency to be $f_* \sim 0.04$ per the discussion from Refs.~\cite{Fukugita:1997bi, Fukugita:2004ee}, which studied the energy budget of the universe in all known forms of matter and radiation---this value is also on the higher end of present day observational constraints~\cite{2019ARA&A..57..227K} and values inferred from stellar mass-halo mass relationships~\cite{2019MNRAS.488.3143B}, assuming the baryon fraction follows the cosmic value.
The form we assumed for $N_*$ also assumes $f_*$ is independent of time, when in reality this quantity changes between different generations of star formation. 
In previous literature, it was believed that star formation peaks at $z\sim2$ and decreases towards earlier redshifts~\cite{Behroozi:2012iw,Madau:2014bja}.
Results from JWST for high-redshift galaxies change this picture, with recent works claiming that the efficiency should rise towards higher redshifts in order to account for the large masses of these galaxies~\cite{2024Natur.635..311X,2025ApJ...988L..35W,2026arXiv260418683P}.
Since the literature on $f_*$ has been evolving rapidly, we will proceed with a fiducial value of $f_* \sim 0.04$ and note that the astrophysical uncertainties on this quantity, particularly towards high redshifts, are large.
Second, we take $f_\mathrm{SN} = 0.0068$.
From the review of cosmic star formation history in Ref.~\cite{Madau:2014bja}, the number of stars that undergo core-collapse supernova per unit mass is expected to be around $k_\mathrm{CC} = 0.0068 \,M_\odot^{-1}$---for solar mass stars, this is approximately one supernova per 150 stars.
Lastly, we take $M_* = 1 \,M_\odot$. 
Note that this mass scale is more appropriate for second or third generation stars forming from enriched gas---the first generation of stars in these early halos will more resemble Pop III stars, which are expected to have larger masses of $M_* \sim 100 \,M_\odot$, smaller values of $f_*$, and values of $f_\mathrm{SN}$ near unity.
These effects compensate one another such that the contribution of this first generation is not more than the generation we consider here.
Moreover, changes in these parameters are subdominant to the effects that we consider next.

Eq.~\eqref{eq:ESN} gives the amount of energy injected by the supernova; however, the energy that is ultimately \textit{deposited} can be much less.
One additional consideration is the supernova coupling efficiency, $f_\mathrm{coup}$, which determines how much of the energy released by a supernova is actually transferred to surrounding gas as kinetic energy.
This number is lower than unity because supernovae can vent through low-density regions, which reduces the net energy that they deposit~\cite{2013MNRAS.431.1337R}.
We will take $f_\mathrm{coup} = 0.1$, which is slightly on the higher end of expected values~\cite{2015MNRAS.451.2757W,2023IAUS..373..199B}.

Another factor to consider is that the energy deposited by supernovae can be dissipated due to radiative cooling by the gas~\cite{1986ApJ...303...39D}.
Radiative cooling is typically parameterized in terms of the cooling function $\Lambda = \frac{1}{n_\mathrm{H}^2} \frac{dE}{dt dV}$, where $n_H$ is the number density of hydrogen.
The cooling function is dependent upon the composition of the gas---in primordial star-forming gas, the cooling function is generally $\Lambda < 10^{-27}$~ergs~cm$^3$~s$^{-1}$~\cite{Galli:1998dh,Maio:2007yf}, while in enriched gas the cooling function is on the order of $\Lambda 
\sim 10^{-23}$~ergs~cm$^3$~s$^{-1}$~\cite{1993ApJS...88..253S}.
Then the energy injected by supernova effectively becomes
\begin{equation}
    E_\mathrm{SN, eff} = f_\mathrm{coup} E_\mathrm{SN} - \Lambda n_\mathrm{H}^2 V t_\mathrm{lifetime}, 
\end{equation}
where $V$ is the volume of the halo and $t_\mathrm{lifetime}$ is the amount of time between when the halo forms and today.
Then, for a given overdensity, the stars and gas within the corresponding halo will become unbound if
\begin{equation}
    E_\mathrm{SN,eff} > E_\mathrm{grav}.
    \label{eq:SN_energy_cond}
\end{equation}

In the limit where $\Lambda \rightarrow 0$, we can rewrite Eq.~\eqref{eq:SN_energy_cond} as a condition on the virial temperature and find that it is equivalent to $T_\mathrm{vir} < \mathcal{O} (10^4)$~K for $f_\mathrm{SN} = 0.0068$.
However, if $\Lambda = 10^{-27}$~ergs~cm$^3$~s$^{-1}$, this is already enough to almost complete diffuse the energy injected by supernova for the halos considered in this work---for the higher value of $\Lambda = 10^{-23}$~ergs~cm$^3$~s$^{-1}$, the effect of supernova is totally negligible.
Hence, the true boundary for which supernova feedback becomes important lies well below $T_\mathrm{vir} \sim 10^4$~K.

\subsection{Collisional systems}
\label{sec:migration}

In a typical galaxy, the stars and dark matter are at a low enough density that they can be treated as a collisionless system~\cite{2008gady.book.....B}.
However, in a denser system such as a globular cluster or early halo, gravitational scatters can become efficient enough for the system to be collisional and the dynamics to become more interesting.

Let us presume that the halos we are interested in are comprised of stars and a smooth dark matter component (i.e. neglecting any granularity in the dark matter structure). 
Since we are interested in stars undergoing dynamical friction, they must be relatively long-lived and therefore cannot be the massive first-generation stars that have mostly gone supernova by today.
Since the lifetime of a star scales with mass approximately as $\sim M^{-2.5}$~\cite{1990sse..book.....K}, a reduction in the star's mass by even one or two orders of magnitude leads to several orders of magnitude increase in the lifetime.
We are therefore motivated to consider lower mass stars in this section and continue to use $M_* = 1 M_\odot$.

Assuming the stellar and dark matter components initially have similar velocities and treating them as gases, equipartition tells us that the more massive component (i.e. the stars) will transfer energy to the lighter component (i.e. the smooth dark matter).
The change in the velocity $\vec{v}$ of the stars due to dynamical friction between stars and dark matter is given by~\cite{1943ApJ....97..255C}
\begin{equation}
    \frac{d \vec{v}}{dt} = - \frac{4 \pi \ln\Lambda G^2 \rho_\mathrm{DM} M_*}{v^3} \left[ \mathrm{erf} (X) - \frac{2 X}{\sqrt{\pi}} e^{-X^2} \right] \vec{v} ,
    \label{eq:dyn_fric}
\end{equation}
where $\ln \Lambda \sim 10$ is the Coulomb logarithm, $G$ is the gravitational constant, $\rho_\mathrm{DM}$ is the local dark matter density, $X = v / (\sqrt{2 \sigma)}$, and $\sigma$ is the dark matter velocity dispersion.

\begin{figure}
    \centering
    \includegraphics[width=0.48\textwidth]{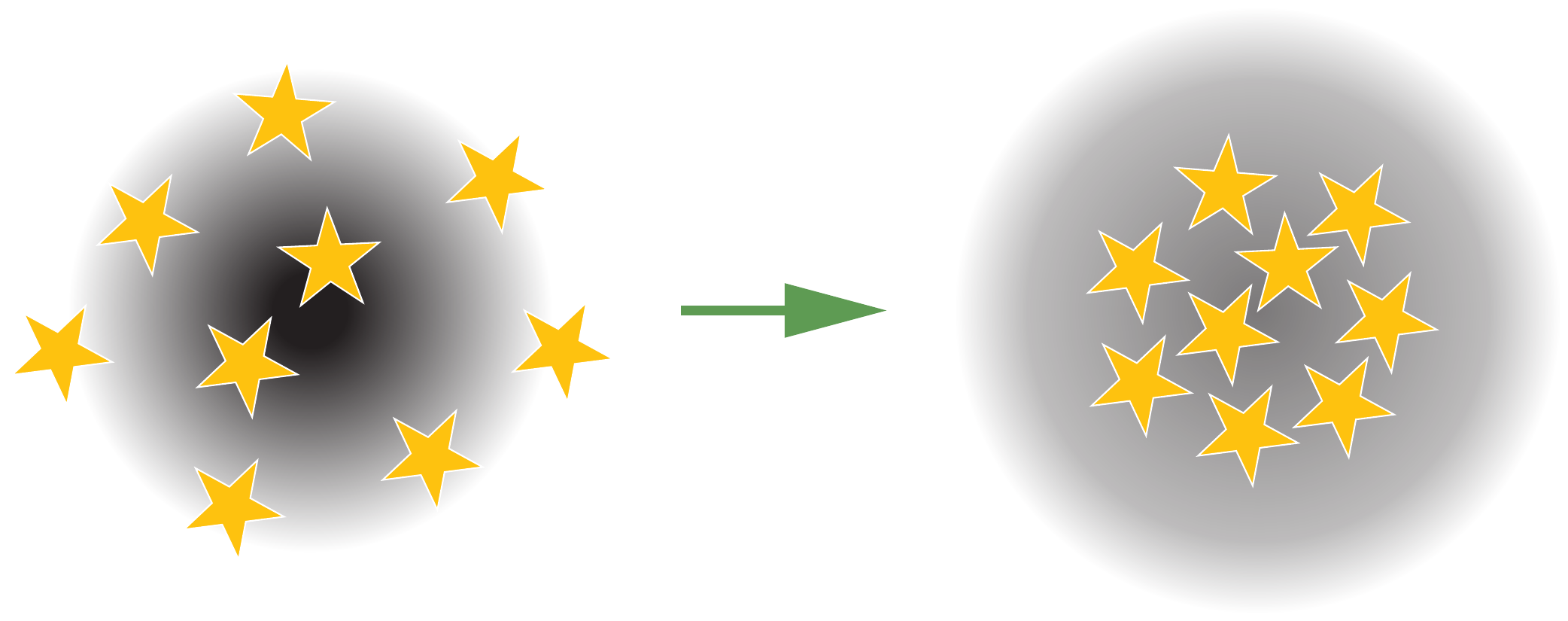}
    \caption{
        A diagram qualitatively illustrating the effect of dynamical friction in a star-forming halo.
        After stars begin forming, they will have some distribution throughout the dark matter halo.
        If the dark matter density is sufficiently large, then the effect of dynamical friction will be significant, and the stars will transfer energy to the dark matter.
        This causes stars to sink towards the center of the halo and cluster, while the dark matter halo expands, although we do not model the latter effect in this work.
    }
    \label{fig:friction}
\end{figure}
As the stars lose energy and angular momentum to dynamical friction, the typical orbital radius will contract and the stellar component will become more compact
\footnote{In later stages of migration, the stars may eventually undergo an ``equipartition instability", or ``gravothermal instability" if they have a mass distribution~\cite{2008gady.book.....B}---we do not consider these phenomena in detail in this work. This is closely related to a phenomenon which occurs in halos comprised of self-interacting dark matter (SIDM), in which the inner core of the halo rapidly increases in density. In the SIDM context, this phenomenon is driven by the self-interactions, not gravitational scatters, and is more commonly referred to as gravothermal core collapse.}.
Fig.~\ref{fig:friction} qualitatively shows how dynamical friction will affect the stellar and dark matter components of a dense halo.
We note that although we only consider a single mass scale here, realistically the stars have a mass distribution and heavier stars will contract more quickly---we comment on the impact of mass distributions in more detail in Section~\ref{sec:caveats}.
Note that in the early stages of the stars contracting, while the dark matter component is still the dominant component of the gravitational potential, the typical velocity of the stars, $v \sim \sqrt{G M_\mathrm{DM} (r) / r}$, is constant or decreasing towards smaller radii for most reasonable dark matter density profiles, where $M_\mathrm{DM} (r)$ is the dark matter mass enclosed in the radius $r$.

As the stars contract and rise to higher densities, they may eventually come to dominate over the dark matter potential and more closely resemble an early star cluster than an early galaxy.
We will use the dynamical friction equation above to determine this threshold between early clusters and early galaxies in two ways.
First, we will use a simple timescale comparison.
Second, we will use toy halo models and solve for the time at which the stellar gravitational potential dominates over the dark matter, and show that the result is not significantly different from the result of the timescale argument.

\subsubsection{Simple timescale estimate}
\label{sec:easy_way}

The timescale for the stars to transfer a significant amount of their kinetic energy is the relaxation timescale, which can be obtained from Eq.~\eqref{eq:dyn_fric} as
\begin{equation}
    t_\mathrm{relax} = \frac{\sigma^3}{\sqrt{2} \pi \ln\Lambda G^2 \rho_\mathrm{DM} M_*} \left[ \mathrm{erf} (1) - \frac{2}{\sqrt{\pi}} e^{-1} \right]^{-1} .
    \label{eq:trelax}
\end{equation}
Then, so long as $t_\mathrm{relax} < t_\mathrm{lifetime}$, it should be possible for the stars to efficiently lose energy and form a dense cluster-like configuration.

\subsubsection{Clustering condition from star orbits}
\label{sec:hard_way}

To show the conditions for clustering more explicitly, we will solve the dynamical friction equation to see when stars come to dominate the gravitational potential.
For simplicity, we consider stars on \textit{circular} orbits within the dark matter halo and do not consider the effect of dynamical friction on the dark matter itself (i.e. the dark matter density profile is held fixed).
Due to dynamical friction, the stars will begin to contract as they orbit and eventually rise to a high enough density to dominate the gravitational potential at the center of the halo---this occurs when
\begin{equation}
    \frac{\rho_*}{\rho_{*,0}} > \frac{\Omega_\mathrm{DM}}{f_* \Omega_b} .
\end{equation}
Here, $\rho_*$ is the mass density of stars, with $\rho_{*,0}$ being the initial value, and $\Omega_\mathrm{DM}$ is the density parameter for dark matter alone.
Assuming that $M_*$ does not change significantly with time, we can recast this as a rough condition on the radius of the stellar component:
\begin{equation}
    f \equiv \frac{r}{r_0} < \left( \frac{f_* \Omega_b}{\Omega_\mathrm{DM}} \right)^{1/3} \sim 0.2 .
    \label{eqn:r_ratio}
\end{equation}
Here we have defined $f$ as the ratio between the stellar component's current radius $r$ to initial radius $r_0$; in the system we are considering, $f < 1$ always.

Now, consider a star within a dark matter halo of some density profile $\rho(r)$. 
Due to dynamical friction, the force exerted on the star is
\begin{gather}
     F = - \frac{4 \pi \ln\Lambda G^2 M_*^2}{v^2 (r)} \rho (r) \left[ \mathrm{erf} (X(r)) - \frac{2 X(r)}{\sqrt{\pi}} e^{-X^2 (r)} \right] , \\ v(r) = \sqrt{ \frac{G}{r} \int_0^r \rho(r') \,d^3 r'} .
\end{gather}
We will assume that the star stays on a circular orbit, but loses angular momentum due to the tangential dynamical friction force.
Denoting the angular momentum per unit mass as $\ell$, then the rate at which the star loses angular momentum is
\begin{equation}
    \frac{d\ell}{dt} = \frac{F r}{M_*} = \left(r \frac{dv}{dr} + v \right) \frac{dr}{dt} .
\end{equation}
The evolution of $r$ over time is therefore
\begin{align}
    \frac{dr}{dt} &= - \frac{4 \pi \ln\Lambda G^2 M_*}{v^2 (r)} r \rho (r) \left[ \mathrm{erf} (X(r)) - \frac{2 X(r)}{\sqrt{\pi}} e^{-X^2 (r)} \right] \n 
    & \qquad \times \left(r \frac{dv(r)}{dr} + v (r) \right)^{-1} .
    \label{eq:r_evol}
\end{align}

In the case where the density profile is an \textit{isothermal sphere},
\begin{equation}
    \rho_\mathrm{iso} (r) = \frac{\sigma^2}{2\pi G r^2} ,
\end{equation}
the dynamics are quite simple.
$\sigma$ is the velocity dispersion of the sphere, so we will take the characteristic circular velocity of objects moving in this halo to be $v_c = \sqrt{2} \sigma$, so that $X = 1$.
We can also write the angular momentum on this circular orbit as $\ell = r v_c$.
In an isothermal sphere, the circular velocity of the star is $v_c$ at all radii, hence
\begin{equation}
    r \frac{dr}{dt} = - \left[ \mathrm{erf} (1) - \frac{2}{\sqrt{\pi}} e^{-1} \right] \ln\Lambda \frac{G M_*}{v_c} .
\end{equation}
Solving this for the evolution of $r$ over time, we have
\begin{equation}
    \frac{r^2}{2} = \frac{r_0^2}{2} - \left[ \mathrm{erf} (1) - \frac{2}{\sqrt{\pi}} e^{-1} \right] \ln\Lambda \frac{G M_*}{v_c} t .
\end{equation}

As stated previously, we want to know when $r = f \times r_0$.
Substituting this in, we find
\begin{equation}
     t_\mathrm{cluster} = \frac{1}{2} \frac{1 - f^2} {\mathrm{erf} (1) - \frac{2}{\sqrt{\pi}} e^{-1}} \frac{r_0^2 v_c}{G M_* \ln\Lambda}.
     \label{eqn:t_cluster}
\end{equation}
Although we assumed an isothermal profile in order to derive this result, we have checked that this result does not depend significantly on the choice of density profile.
See Appendix~\ref{app:dens_prof} for details on this cross-check.
Furthermore, comparing this expression to the relaxation time, we find that
\begin{equation}
    t_\mathrm{cluster} = \frac{1 - f^2}{2} t_\mathrm{relax} .
\end{equation}
Since $0 < f < 1$, then $t_\mathrm{cluster} \lesssim t_\mathrm{relax}$ and the argument in Section~\ref{sec:easy_way} using only the relaxation time is a sufficient condition for clustering to occur. 

\subsubsection{Caveats and limitations}
\label{sec:caveats}

Besides employing a simulation, there are several ways in which this calculation could be improved.
For example, here we have not considered the response of the dark matter to the dynamical friction.
According to this picture, after the stars cluster at the center of the potential, the dark matter halo should be more diffuse and extended due to the kinetic energy that has been transferred to it---although the dynamics of the stars will no longer trace the dark matter profile, it would be interesting to model the expansion of the dark matter halo and explore other possible observables. 
Mass-to-light ratios of globular clusters constrain the content of dark matter within the cluster to be $\lesssim 20\%$~\cite{2020PASA...37...46B}; however, attempts to constrain scenarios where the globular clusters are surrounded by a halo of dark matter yield ambiguous results~\cite{,2021ApJ...922..104C,Garani:2023esk}.
This scenario of forming star clusters by dynamical friction is also closely related to the simulation studies of Ref.~\cite{Errani:2026hsq}, which demonstrated clustering by dynamical friction within ultrafaint galaxies, although they did not consider halos with enhanced initial densities.

As another improvement, we assumed circular orbits for simplicity; however the dynamical friction timescale decreases for eccentric orbits~\cite{1999ApJ...515...50V}.
These effects are unlikely to change the condition for cluster formation by orders of magnitude, and we therefore defer a study including these corrections to future work.

In addition, we have assumed a single mass scale for the stars in these enhanced density halos.
However, if there is a significant spread in the initial mass function of the high redshift stars, then higher mass stars will contract more quickly than the low mass stars, resulting in mass segregation.
If this process is very efficient, then the fate of this system depends on whether the supernova rate or merger rate is dominant for the high-mass stars clustered at the halo's core.
Merging of high mass stars at the center of the cluster may lead to the formation of a central intermediate-mass black hole~\cite{2004Natur.428..724P,2010ARA&A..48..431P}, provided that collisions occur more rapidly than the infalling stars can go supernova.
This provides a mechanism for forming SMBH seeds independent of the direct-collapse that leads to NQPBHs at high redshifts.
On the other hand, if supernovae are efficient, then the stars will explode faster than they can assemble into a large central object.
Estimating the effect of mass segregation requires knowing the initial mass function of these high redshift stars, which is highly uncertain and we consider to be outside of the scope of present work.

Another consideration that we neglect in this work is substructure within the halos considered above. 
As discussed in Ref.~\cite{Graham:2024hah}, a related scenario to the one considered in this work involves constraining effects from dynamical heating due to dark matter substructure on stars within ultrafaint dwarf galaxies.
In that work, since the dark matter clumps are significantly larger than the stars within the dwarf galaxy, the primary effect is for the dark matter substructure to transfer energy to the stellar component and therefore expand the stellar radius.
This granularity may also apply to our scenario and counter the clustering of stars considered here.
On the other hand, the dynamical friction experienced by stars themselves may reduce the expansion of the stellar radii, expand the host dark matter halos and render them more susceptible to disruption, possibly relaxing the constraints presented in Ref.~\cite{Graham:2024hah}.
We leave a quantitative assessment of this effect to future work that includes the feedback of dynamical friction on the dark matter halos.

\section{Results}
\label{sec:results}

\begin{figure*}
    \centering
    \includegraphics[width=0.75\textwidth]{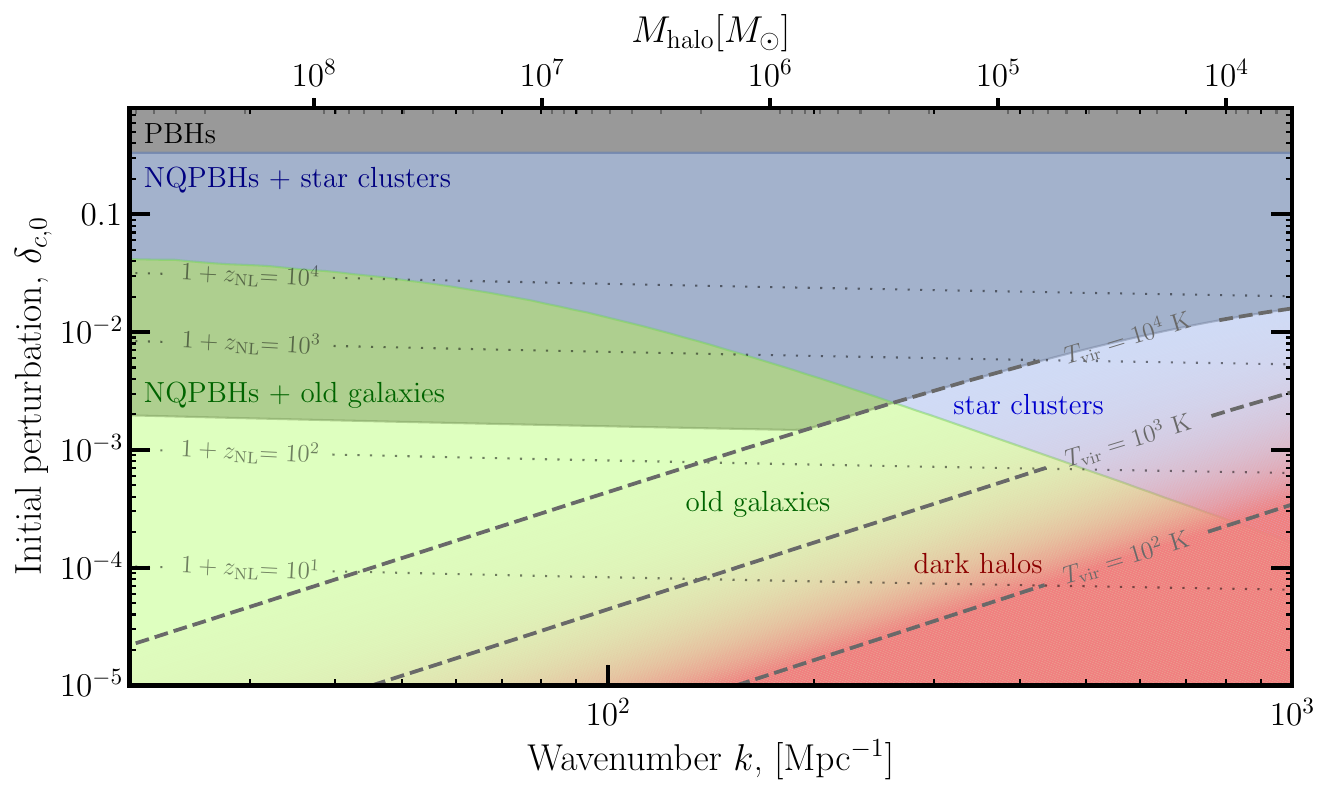}
    \caption{
        A summary of the stellar systems that may form, given the initial density perturbation and wavenumber $k$ (bottom axis) or corresponding halo mass $M_\mathrm{halo}$ (top axis). 
        The regions are divided into dark halos, where stars are either unable to form or remain bound due to the effects of supernova (red); old galaxies, which form when stars are able to form and remain bound, but unable to cluster (green); and star clusters, which form when dynamical friction is efficient (blue).
        The fuzziness of the dark halos contour represents the variation in this curve due to several uncertainties in the level of supernovae feedback.
        We also show the parameter space where it is possible to form NQPBHs (gray) and PBHs (dark gray).
    }
    \label{fig:summary}
\end{figure*}

Having established a few of the relevant halo properties, we can now discuss the requisite conditions for a particular overdensity to form different stellar systems.
Fig.~\ref{fig:summary} summarizes the types of systems that can form given the amplitude of the initial density fluctuation and the wavenumber or corresponding halo mass.
For reference, the parameter space where overdensities become PBHs is shown in dark gray---in this region, \textit{all} overdensities become PBHs.
The parameter space where it is possible to form NQPBHs is shown in lighter gray~\cite{Qin:2025ymc}.
Only the fraction of overdensities that have sufficiently low angular momentum and dark matter-baryon relative velocities form NQPBHs; otherwise, collapse of the baryons into compact objects is delayed until sufficient molecular hydrogen forms to allow the gas to fragment.
Ref.~\cite{Qin:2025ymc} found that these effects could reduce the number of halos capable of forming NQPBHs by a factor of $\mathcal{O} (10)$.
We stress that there therefore exist halos within the NQPBH parameter space that will fail to form NQPBHs and instead result in other astrophysical objects, as detailed below.
The contours for $T_\mathrm{vir} = 10^2, 10^3,$ and $10^4$\,K are also shown in Fig.~\ref{fig:summary}. 

Based on the conditions derived in Section~\ref{sec:stellar}, we divide the remaining overdensities into three key regions:
\begin{itemize}
    \item \textbf{Dark halos} (red): 
    In this region, the virial temperature for baryons falling into the halo's gravitational potential is less than $10^2-10^3$\,K, which is below the molecular cooling threshold for gas to be able to form stars.
    A halo may also be starless if $E_\mathrm{SN,eff} > E_\mathrm{grav}$ and supernovae deposit energy into the baryons fast enough to unbind gas/stars by today.
    Hence, in such objects, we expect there to be little luminous matter left by today.
    Under our least optimistic assumption, i.e. when supernova feedback is strongest, we find that this condition is equivalent to $T_\mathrm{vir} \lesssim 10^4$~K.
    However, if the composition of gas in these early halos is such that radiative cooling is in fact non-neglible, the effect can greatly reduce the effect of supernova.
    Given these uncertainties, we represent the dark halo contour with a gradient between $10^2\,\mathrm{K} < T_\mathrm{vir} < 10^4\,\mathrm{K}$ as opposed to a sharp contour.

    \item \textbf{Old galaxies} (green):
    In this region, $E_\mathrm{SN,eff} < E_\mathrm{grav}$, meaning the luminous matter in the halo survives the effect of supernovae, and $t_\mathrm{lifetime} < t_\mathrm{cluster}$, meaning dynamical friction is not significant.
    In this case, the stars will resemble a galaxy that formed at exceedingly high redshifts.

    \item \textbf{Star clusters} (blue):
    In this region, the halos are capable of forming and retaining stars, and the effect of dynamical friction is also significant.
    The resulting stellar system would therefore be more compact than a high-redshift galaxy.
    Hence, the stellar systems that form from such overdensities will more closely resemble star clusters, which are bound under the stars' self-gravity, than galaxies, which are bound by dark matter.
\end{itemize}

The last region is interesting as a possible pathway to form globular clusters, whose formation is still not well understood.
Our results show that overdensities forming with initial amplitudes greater than $\delta_{c,0} \gtrsim 4 \times 10^{-3}$ at a wavenumber of $k \sim 200$~Mpc$^{-1}$ can form star clusters with stellar masses of approximately $\frac{\Omega_b}{\Omega_m} M_\mathrm{halo} f_* \sim 5000 \, M_\odot$, which is at the low end of observed globular cluster masses~\cite{2018ARA&A..56...83B}, although the cluster mass may be higher if $f_*$ is larger than the value of 0.04 assumed here---this uncertainty was discussed in Section~\ref{sec:supernovae}.
Larger clusters can also result if the overdensity forms from a lower wavenumber mode and higher initial value.
This mechanism also links the formation of star clusters to dwarf galaxies, since both would be formed from relatively small dark matter halos, with efficient dynamical friction leading to the former. 
A link between the origins of these two classes of objects is an interesting consequence, given suggestions that the populations are not entirely separate~\cite{2026arXiv260217652C,Errani:2026hsq}.

A compelling formation pathway for globular clusters must also explain their varied chemical abundances, which indicate the presence of ``multiple populations" within the cluster~\cite{2018ARA&A..56...83B,2022Univ....8..359M}.
Given that the overdensities considered in our scenario form earlier than even standard Pop III stars under standard cosmological assumptions, they have unusually deep potential wells.
As discussed in Section~\ref{sec:supernovae}, these massive and early stars will likely have all gone supernovae by today, enriching the surrounding gas and changing the composition of subsequent generations of stars.
Such a scenario is in line with a common hypothesis for globular cluster formation, which states for stars forming in a deep potential well, the ejecta from a first generation of stars can be retained and undergo later generations of star formation~\cite{2018ARA&A..56...83B,2019A&ARv..27....8G}, and also the fact that there is no evidence for Pop III stars in globular clusters today~\cite{1981ApJ...248..606B}.
We also note that recent work suggests that the magnesium and aluminum abundances in some globular clusters may be explained by very hot hydrogen burning in supermassive stars~\cite{2026arXiv260909271K}, which are more likely to form in the early and dense halos considered here~\cite{2024PASJ...76..850I}.
We defer a detailed exploration of the chemical abundances in these early stellar systems to future study.

\section{Further predictions for early structure}
\label{sec:predictions}

As a result of this and previous work~\cite{Qin:2025ymc}, we have shown that both SMBHs and some star clusters may have their origins in enhanced density fluctuations.
Given a particular power spectrum with small length scale enhancements, we discuss here steps to predict further observables, such as the abundance of star clusters, the contributions of early stars to reionization, as well as changes in dark matter substructure which may be detectable with stellar stream surveys~\cite{Bonaca:2024dgc}.

\begin{figure}
    \centering
    \includegraphics[width=\columnwidth]{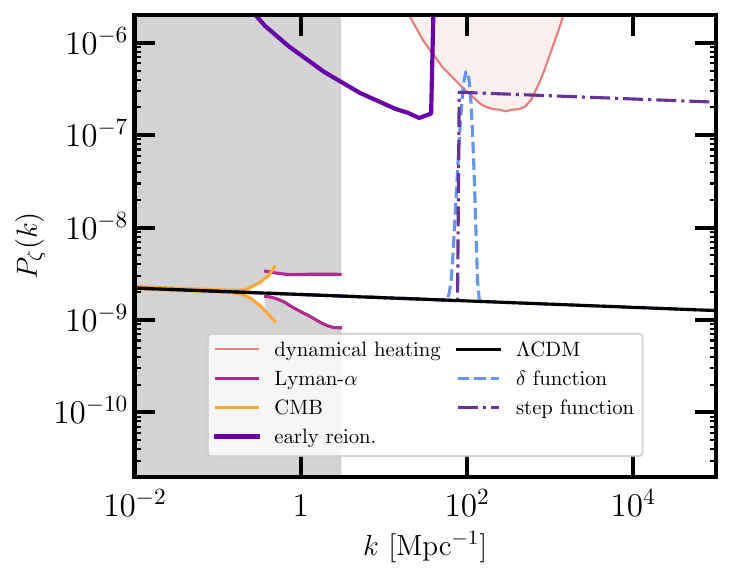}
    \caption{
    	The primordial curvature power spectrum.
        The grey contours show constraints from CMB anisotropies~\cite{Planck:2018jri} and Lyman-$\alpha$ forest measurements~\cite{2011MNRAS.413.1717B}.
        The pink contour shows the dynamical heating constraints from Ref.~\cite{Graham:2024hah} and the purple curve shows the estimated constraint due to early reionization from Ref.~\cite{Qin:2025ymc}.
        We note that in this wavenumber range, there also exist limits from strong lensing~\cite{Gilman:2021gkj} and dwarf galaxy properties~\cite{Esteban:2023xpk,Balaji:2024wkv,Dekker:2024nkb}---while these observables are applicable to the enhancements we consider, we do not display the constraints here since these studies assume different forms for the enhancements. 
        The black curve shows the standard slow-roll power spectrum with parameters taken from Planck 2018~\cite{Planck:2018vyg}, labeled as ``$\Lambda$CDM".
        The dashed curve shows our fiducial spectrum enhanced with a narrow peak or approximate $\delta$-function, while the dot-dashed curve shows the fiducial step function enhancement.
    }
    \label{fig:pspec_fid}
\end{figure}
For this work, we will take the same fiducial primordial curvature power spectra as in Ref.~\cite{Qin:2025ymc}.
Starting with the best fit slow-roll power spectrum from Planck 2018~\cite{Planck:2018vyg}, we add either a narrow Gaussian peak (i.e. an approximation of a $\delta$-function) at $k = 100 \,\mathrm{Mpc}^{-1}$ with a width of $\Delta k = 10 \,\mathrm{Mpc}^{-1}$ that enhances the power by a factor of $\sim 300$, or a step function that enhances power for $k > 80 \,\mathrm{Mpc}^{-1}$ by a factor of 180.
The resulting power spectra are shown in Fig.~\ref{fig:pspec_fid}.
While these models are idealized and not the result of a realistic underlying mechanism to produce enhancements in matter power, we nevertheless find them to be useful benchmarks for how enhancements at specific scales will impact astrophysical observables.

\subsection{Abundance of clusters}

In this section, we will discuss how to estimate the initial abundance of star clusters given a power spectrum.
When calculating the mass function for e.g. dark matter halos, one often uses the Press-Schechter halo mass function, which we denote by $\left( \frac{dn}{dM} \right)_\mathrm{PS}$. 
The Press-Schechter formalism requires calculating when linearly-evolving density perturbations exceed some constant critical threshold, $\delta_c \approx 1.69$, where this threshold is the same across all redshifts.

However, from examining Fig.~\ref{fig:summary}, the contour above which density perturbations can form star clusters is decidedly not constant as a function of $k$.
Critical collapse thresholds that are not constant have been considered as improvements upon Press-Schechter---for example, motivated by considering ellipsoidal collapse, the Sheth-Tormen approximation uses a threshold that is a function of the mass variance $\sigma$~\cite{2001MNRAS.323....1S}.
In the case of star clusters, the collapse threshold is more straightforward to understand as a function of the corresponding wavenumber $k$.
Within the excursion set formalism, one can calculate the mass function of collapsed objects with a nontrivial threshold~\cite{Zentner:2006vw}, e.g. using the algorithms described by Ref.~\cite{2006ApJ...641..641Z} and ~\cite{2013MNRAS.428.1774B}.
These methods are also implemented in galaxy formation codes such as \texttt{Galacticus}~\cite{Benson:2010kx}; however, we find that running this algorithm down to relatively small halo masses of $\lesssim 10^6 \, M_\odot$ is too time-prohibitive for this application.

\begin{figure}
    \centering
    \includegraphics[width=\columnwidth]{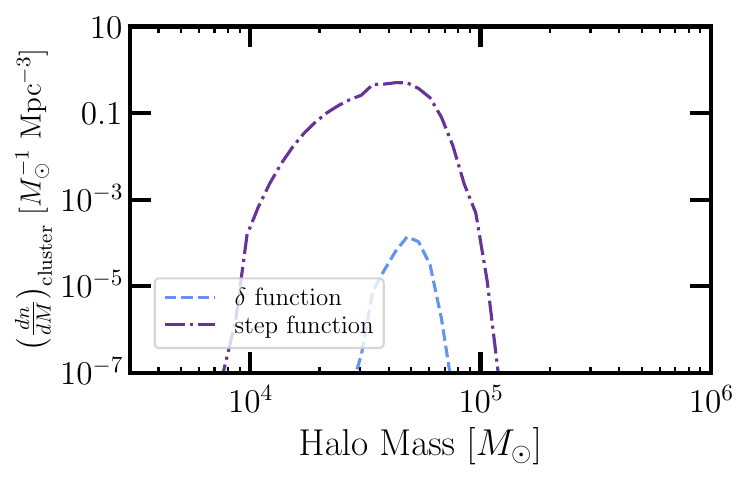}
    \caption{
    	The initial cluster mass function, as calculated using Eq.~\eqref{eqn:dndm_cluster} for the fiducial power spectra shown in Fig.~\ref{fig:pspec_fid}.
        The mass shown on the $x$-axis is the total initial mass of the cluster, including the initial dark matter halo, and not just the stellar mass.
        As described in the text, the calculation of this mass function does not account for the destruction or merging of star clusters.
    }
    \label{fig:clusters}
\end{figure}
Therefore, we will instead approximate the \textit{initial cluster mass function} as follows.
\begin{align}
    \left( \frac{dn}{dM} \right)_\mathrm{cluster} &= \int dz \, \left( \frac{dn}{dM dz} \right)_\mathrm{new} \n
    &\qquad \times \Theta [M - M_\mathrm{mol} (z)] \Theta [M_\mathrm{th} (z) - M]
    \label{eqn:dndm_cluster}
\end{align}
Here, $\Theta (x)$ represents the Heaviside step function.
In this expression, at a given redshift, $M_\mathrm{mol} (z)$ is the mass threshold above which a halo can form astrophysical objects and $M_\mathrm{th} (z)$ is mass threshold below which a halo can form a star cluster.
These thresholds can be determined from Fig.~\ref{fig:summary} by taking the intersection of the $z_\mathrm{NL} = z$ contour with the $T_\mathrm{vir} = 10^3\,\mathrm{K}$ contour or the $t_\mathrm{cluster} = t_\mathrm{lifetime}$ contour, respectively.
The quantity $\left( \frac{dn}{dM dz} \right)_\mathrm{new}$ represents the rate at which new halos form; however, due to the cloud-in-cloud problem, smaller halos can eventually be subsumed into larger halos, leading to $\frac{d}{dt} \left( \frac{dn}{dM} \right)_\mathrm{PS} < 0$ at small masses and late redshifts.
Since clusters may survive within halos which are themselves \textit{not} clusters, including negative values of $\frac{d}{dt} \left( \frac{dn}{dM} \right)_\mathrm{PS}$ could lead to an underestimate in cluster abundance.
Hence, we take the rate at which new halos form to be
\begin{align}
    \left( \frac{dn}{dM dz} \right)_\mathrm{new} &= 
    \begin{cases}
        \frac{d}{dz} \left( \frac{dn}{dM} \right)_\mathrm{PS} , & \frac{d}{dz} \left( \frac{dn}{dM} \right)_\mathrm{PS} < 0, \\
        0, & \frac{d}{dz} \left( \frac{dn}{dM} \right)_\mathrm{PS} \geq 0 .
    \end{cases}
\end{align}
In other words, to get the cluster abundance, we take the new halos formed at each redshift and determine how many of them are within the right mass range to become a star cluster. 

We emphasize that for all initial power spectra, including the $\Lambda$CDM case, we are \textit{only} considering clusters formed by this dynamical friction mechanism.
Moreover, Eq.~\eqref{eqn:dndm_cluster} does not fully take into account the fact that clusters can merge, their masses can evolve, or that clusters can be destroyed (which we discuss further below).
In particular, since clusters may be able to merge and thus grow their mass, this initial cluster mass function should not be considered a strict upper bound on the abundance of clusters at any particular mass.

Using the fiducial power spectra described previously, Fig.~\ref{fig:clusters} shows our estimate for the initial cluster mass function.
The number of clusters that would be produced within $\Lambda$CDM by this mechanism is so small that it lies many orders of magnitude below the bottom edge of the figure.
Across all halo masses, and for both types of power spectrum enhancements, the abundance of star clusters is enhanced relative to $\Lambda$CDM.

For the narrowly peaked enhancement, we see clusters forming from halo masses between $\sim 3 \times 10^4 - 7 \times 10^4 \,M_\odot$, corresponding to stellar masses of $200-500 \,M_\odot$, and the largest enhancement is at a halo mass of $4 \times 10^4 \,M_\odot$.
Integrating the initial cluster mass function across all masses, we obtain a number density of $n_\mathrm{cluster} = 2.0$~Mpc$^{-3}$ in the narrow-peak enhanced cosmology.
The Milky Way would therefore initially capture about $50$ of these clusters for the narrow-peak case.
Compare this to the $\Lambda$CDM cosmology, where the resulting density is $n_\mathrm{cluster} = 7 \times 10^{-30}$~Mpc$^{-3}$ and yields no clusters within the Milky Way due to this mechanism.
Hence, this benchmark model forms significantly more clusters by dynamical friction; however, the resulting clusters masses are fewer and smaller in mass than observed clusters in the Milky Way, of which there are about 150 clusters spanning stellar masses of $\mathcal{O} (10^3-10^6) \, M_\odot$~\cite{1996AJ....112.1487H}.

In comparison, the step function enhancement forms clusters between $8000 - 10^5 \,M_\odot$, corresponding to stellar masses of $60-800 \,M_\odot$, and shows greater abundances of clusters across all masses compared to the narrowly peaked case---this is due to the fact that in the step-function case, increases in the mass variance continue even down to very small mass scales.
Integrating the initial cluster mass function across all masses, we obtain a number density of $n_\mathrm{cluster} = 1.5 \times 10^4$~Mpc$^{-3}$ in the step-function enhanced cosmology.
The Milky Way would therefore initially capture about $4.1 \times 10^5$ clusters in the step-function case---far more than the observed value, although not all of them may survive until today.

For both of our benchmark models, the stellar masses of the initial clusters are only $\mathcal{O} (100) \,M_\odot$.
For the narrow peak benchmark, there is not enough stellar mass in the initial clusters to assemble into the clusters observed in the Milky Way today.
On the other hand, the initial clusters resulting from the step function benchmark contain about $1.4 \times 10^8 \,M_\odot$ worth of stars, which is slightly more than what is seen in Milky Way clusters.
Hence, if some these star clusters can merge many times, then they may assemble into much larger clusters that more closely match the properties of Milky Way clusters---they are moreover likely to survive until today given that the stars have already migrated into relatively high-density configurations and that they can be protected from tidal effects (at least temporarily) by their surrounding dark matter halo.
Even if mergers happen early, the migration described in Section~\ref{sec:migration} can still proceed inside of the resulting dense halos---likely, simulations will be required to explore all the possible outcomes for these initial clusters.
The initial cluster masses are also dependent on the type of enhancement; as can be seen from Fig.~\ref{fig:summary}, enhancements of both large amplitude and smaller wavenumber can yield more massive clusters.
While larger amplitude/smaller wavenumber enhancements could result in power spectra that are in conflict with the constraints discussed in Fig.~\ref{fig:pspec_fid}, these tensions can also be alleviated if the enhancements are non-Gaussian.
We leave explorations of these effects on cluster mass to future work.

Moreover, these results show that the initial abundance of clusters is highly sensitive to whether the power spectrum enhancement is broad or narrow.
As stated previously, there are only about 150 known globular clusters in the Milky Way~\cite{1996AJ....112.1487H}.
Even if the initial number of clusters is much larger than this, the number of \textit{surviving} star clusters today is likely to be much smaller, since it is known that many processes (evaporation, tidal shocks, dynamical friction, etc.) contribute to their destruction~\cite{1977MNRAS.181P..37F,Gnedin:1996sa} and can reduce the mass function by orders of magnitude, particularly for smaller mass clusters~\cite{Fall:2001ii,2012MNRAS.421.1927K,2015MNRAS.454.1658K}.
In particular, due to the mass-dependent destruction rates, different shapes for the initial cluster mass function can yield similar mass functions today~\cite{Fall:2001ii}.
Hence, a cleaner test of globular cluster formation history would be to look towards high redshifts, before these destruction processes erase too much information about the initial cluster mass function.
If one can obtain more reliable inferences of the initial mass function and in particular if one can find evidence for these systems transitioning from dark matter dominated systems to star-dominated systems, such a result might point towards star clusters resulting from enhanced small scale structure.
Relatedly, there exist proposals that the relatively high-redshift population of LRDs are globular clusters in formation~\cite{Chisholm:2026cbm}.
We leave a more detailed study of the evolution of the cluster mass function to future work.

\subsection{Early star formation and reionization}

As has been discussed throughout this manuscript, the early structure formation resulting from enhanced matter power can push the onset of Pop III star formation to significantly higher redshifts than previously expected~\cite{2015ApJ...814...18H,2024PASJ...76..850I}.
Hence, not only would these stellar systems be observed out to higher redshifts, but systems that survive to the present day would have more time to undergo multiple generations of star formation.

Another consequence is that the early stars, binary systems, 
and black holes resulting from enhanced structure would act as high-redshift sources of ionizing ultraviolet radiation.
Hence, when calculating the global free electron fraction of the universe, $x_e = n_e / n_H$, where $n_e$ and $n_H$ respectively denote the number density of free electrons and total hydrogen nuclei, one should include an additional contribution due to these sources, although accurately modeling astrophysical contributions to reionization is still an active area of research~\cite{McQuinn:2015icp}.
Ref.~\cite{Qin:2025ymc} estimated a constraint on the primordial curvature power spectrum from early reionization using the model of Ref.~\cite{Furlanetto:2004nh}, which assumes that $x_e$ is proportional to the fraction of matter in collapsed structures---the constraint is reproduced in Fig.~\ref{fig:pspec_fid} in purple.
Because of the simplicity of the reionization model, that work assumed a low ionization efficiency and allowed for complete reionization as early as $1+z = 20$, in order to ensure the estimate was not overly stringent.
Hence, a more realistic model for the additional contributions to $x_e$ will likely lead to a stronger constraint.
We note that in the case where feedback mechanisms are able to shut off star formation or black hole accretion shortly after their onset, the change to $x_e$ may not be a monotonic function of redshift, but rather give a peak at high redshifts not unlike the ``flash ionization" scenarios considered in Refs.~\cite{Tan:2025obi,Aggarwal:2026ogm}.

While the exact shape and amplitude of the additional contribution to the global ionization history depends on the modeling of ionizing sources, the prediction of enhanced ionization is robust.
However, further study of this phenomenon would significantly expand the scope of present work and, as such, we leave a detailed study of the impact on reionization to future work.

\subsection{Abundance of dark matter subhalos}

\begin{figure}
    \centering
    \includegraphics[width=\columnwidth]{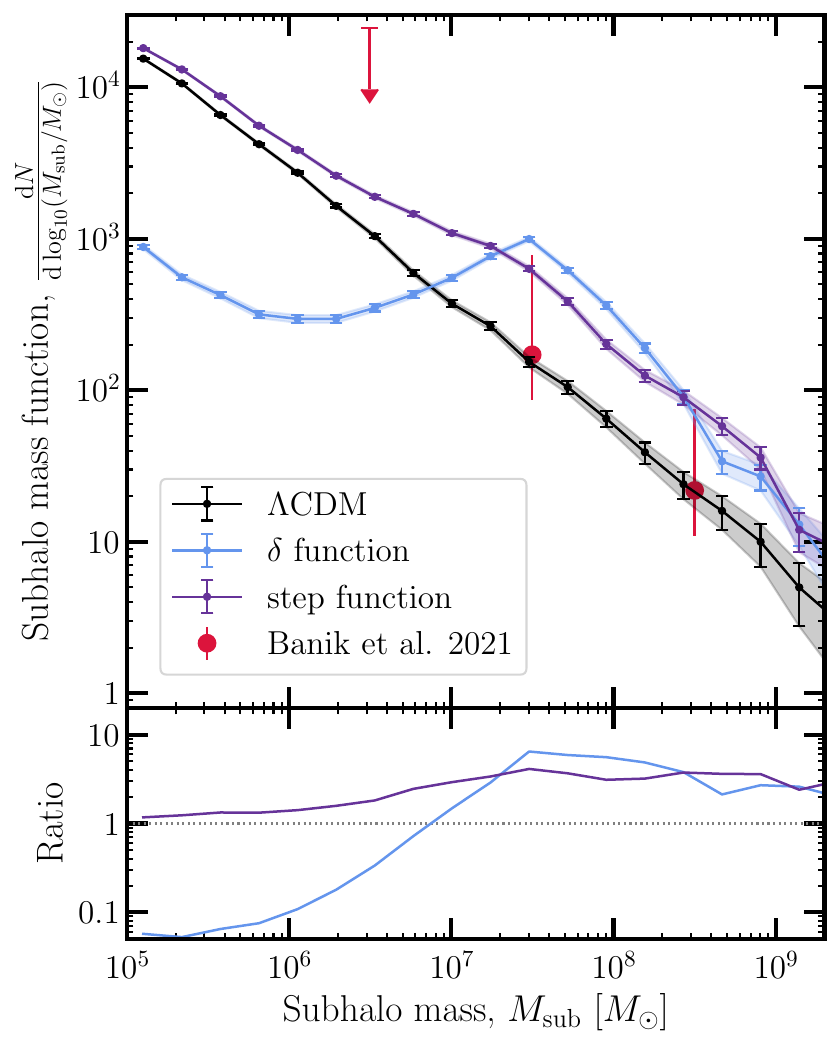}
    \caption{
    	Results from \texttt{Galacticus} for the substructure of a $10^{12} \,M_\odot$ halo using different initial power spectra.
        \textit{Top:} The subhalo mass function, binned by the present-day mass of the subhalos.
        Black shows the predictions from $\Lambda$CDM, blue shows the results for our fiducial narrow-peak enhancements, and purple shows the result of the fiducial step-function enhancement.
        The error bars correspond to $1\sigma$ Poisson uncertainties.
        For reference, we also show measurements of the subhalo mass function from Ref.~\cite{Banik:2019smi}.
        \textit{Bottom:} Ratio of the subhalo mass function in the enhanced cases compared to the $\Lambda$CDM case.
        For the narrow peak, the abundance of subhalos is suppressed below $M_\mathrm{sub} \sim 10^7 \,M_\odot$ and enhanced above this scale by as much as $6.5$ compared to $\Lambda$CDM.
        For the step-function, there is no suppression across the range of mass scales considered here.
    }
    \label{fig:shmf}
\end{figure}

\begin{figure}
    \centering
    \includegraphics[width=\columnwidth]{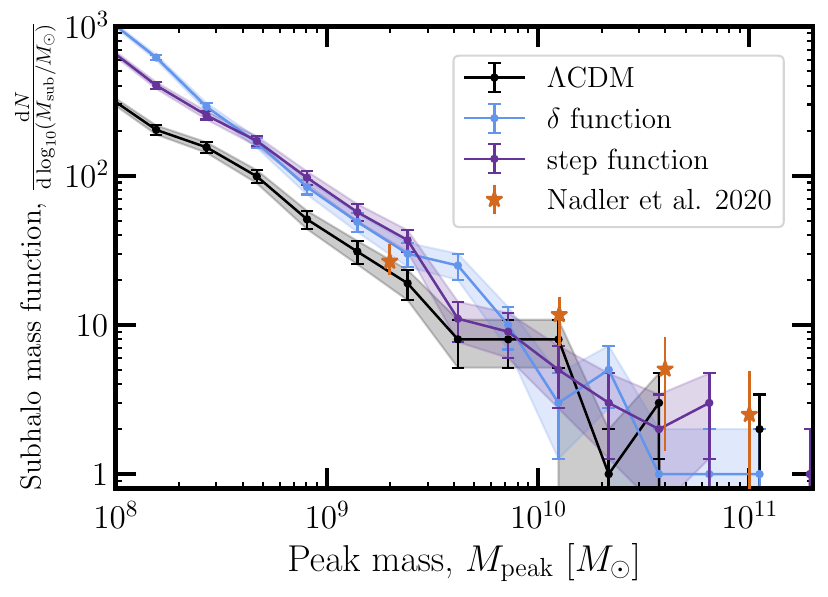}
    \caption{
    	The mass function of the same subhalos as in Fig.~\ref{fig:shmf}, but binned by peak mass instead of present-day mass.
        For comparison, we show measurements from the Milky Way satellite census of Ref.~\cite{DES:2019ltu}.
    }
    \label{fig:shmf_Mpeak}
\end{figure}

In addition to an increase in the number of luminous objects, enhancements in power should lead to increased dark matter substructure (see e.g. Ref.~\cite{Nadler:2025crd} for a more detailed study on the impact of enhancements in power on substructure).
Close encounters between dark matter subhalos and stellar streams can leave gaps and spurs in the stream, as well as features in the phase space density; hence, future stellar stream surveys will be sensitive probes of dark matter substructure and may be able to detect or rule out enhanced small-scale power~\cite{Bonaca:2024dgc,2026arXiv260618332T}.

Using \texttt{Galacticus}~\cite{Benson:2010kx}, we calculate the subhalo mass function of a Milky Way-like halo (i.e. $M_\mathrm{halo} = 10^{12} \,M_\odot$) with a mass resolution of $10^6 \,M_\odot$ on the same fiducial power spectra.
The subhalos are assumed to have a Navarro-Frenk-White (NFW) density profile,
\begin{equation}
    \rho_{\rm NFW}(r) = \frac{\rho_s}{\frac{r}{r_s}\left(1 + \frac{r}{r_s}\right)^2} ,
    \label{eq:NFW}
\end{equation}
where $r_s$ is the scale radius of the subhalo and the scale density $\rho_s$ is set by the subhalo mass.

Fig.~\ref{fig:shmf} shows the resulting subhalo mass functions in terms of the present-day masses of the subhalos; the bottom panel shows the ratio of the mass function in the enhanced cases relative to the $\Lambda$CDM case.
The error bars correspond to $1\sigma$ Poisson uncertainties.
Compared to the $\Lambda$CDM prediction, the narrow-peak enhanced power spectrum leads to suppression of structure at subhalo masses less than $10^7 \,M_\odot$, and enhancements above this scale.
The suppression is caused by the fact that $10^6 \,M_\odot$ halos form much earlier in the enhanced power cosmology than $\Lambda$CDM, and therefore have more time to either merge, be accreted, or destroyed.
The maximum enhancement is 6.5 times larger than the $\Lambda$CDM prediction at a subhalo mass of $3 \times 10^7 \,M_\odot$.
For the step-function enhancement, the subhalo mass function is always enhanced by a factor of a few or comparable to the $\Lambda$CDM mass function, with the largest enhancements at intermediate masses between $6 \times 10^6 \,M_\odot$ to $10^9 \,M_\odot$.
As a validation check, our results are broadly in agreement with Ref~\cite{Nadler:2025crd}, which also found a factor of $\mathcal{O}(1)$ enhancement in subhalo abundance at certain masses, albeit using a smaller power spectrum enhancement and different shapes.
In Fig.~\ref{fig:shmf}, we also show the subhalo mass function measurements from Ref.~\cite{Banik:2019smi}, which were inferred from perturbations in the GD-1 and Pal 5 stellar streams.
For the fiducial narrow-peak enhancement that we consider, the subhalo mass function yielded by \texttt{Galacticus} is mostly within agreement of the measurements from Ref.~\cite{Banik:2019smi}.
The narrow-peak enhancement is slightly in tension with the measurement at $3 \times 10^7 M_\odot$, but within bounds for the other mass bins.

There also exist measurements of Milky Way satellite abundances using data from the Dark Energy Survey (DES) and Pan-STARRS1~\cite{DES:2019ltu}.
The results of that work are presented not in terms of the present-day subhalo mass, but the largest mass attained by the subhalo at any point in its lifetime, i.e. the peak mass $M_\mathrm{peak}$.
The closest quantity that \texttt{Galacticus} computes to the peak mass is the subhalo's mass at the time of infall---hence, for the \texttt{Galacticus} outputs, we use the infall mass as a proxy for $M_\mathrm{peak}$.
In Fig.~\ref{fig:shmf_Mpeak}, we show the subhalo peak mass function from \texttt{Galacticus} against the measurements of Ref.~\cite{DES:2019ltu}.
At these larger masses, the differences in the mass function between the $\Lambda$CDM and enhanced cosmologies are smaller and have larger relative error; therefore, while it is necessary to check that our predicted mass functions are consistent with Milky Way satellites, the mass function at this scale does not distinguish small scale enhancements as cleanly as at smaller masses.
In the mass bin at $M_\mathrm{peak} = 2 \times 10^9 M_\odot$, the $\Lambda$CDM and narrow-peak enhancement mass functions are in agreement with the data, while the step-function case slightly overshoots the data point.
For subhalos with masses greater than $10^{10} M_\odot$, the merger tree becomes very sparse and the mass function is therefore quite noisy, but broadly in agreement with Ref.~\cite{DES:2019ltu}.

\begin{figure}
    \centering
    \includegraphics[width=\columnwidth]{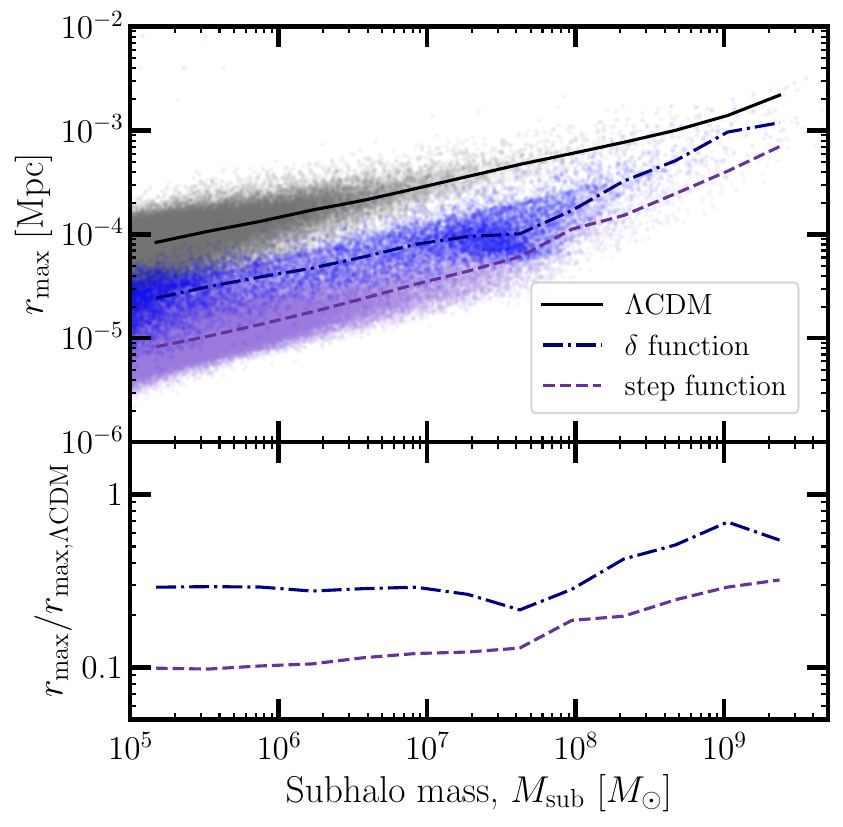}
    \caption{
    	Radius of maximum circular velocity for each subhalo compared to mass for the same halo as in Fig.~\ref{fig:shmf}.
        We show both a scatter plot and curves for the median $r_\mathrm{max}$ binned by mass.
        Because the subhalos in the enhanced power cosmologies form earlier, they are more compact than subhalos within $\Lambda$CDM.
        This can enhance their detectability within e.g. stellar stream surveys.
    }
    \label{fig:rmax}
\end{figure}

Because halos that form earlier have a higher virial density, we expect them to be more compact today compared to their $\Lambda$CDM counterparts.
A common way to parameterize this is in terms of the halo mass-concentration relation; however, given the ambiguities in defining concentration for subhalos due to tidal stripping~\cite{Moline:2016pbm}, we will instead characterize the relative compactness of subhalos using $r_\mathrm{max}$, the radius of their maximum circular velocity.
Fig.~\ref{fig:rmax} shows $r_\mathrm{max}$ for the subhalos versus their mass. 
The scatter plots show $r_\mathrm{max}$ and the mass of every halo; the curves show $r_\mathrm{max}$ binned by mass.
Comparing $r_\mathrm{max}$ across the three different initial power spectra, we see that the radii of subhalos in the enhanced cosmologies are indeed systematically smaller than their $\Lambda$CDM counterparts, due to their increased densities from forming at high redshifts.
Since more compact subhalos are able to deliver larger velocity kicks to stellar streams~\cite{2015MNRAS.450.1136E,2015MNRAS.454.3542E}, then subhalos resulting from enhanced power spectra will be easier to detect in stellar stream surveys, even at the lowest masses of $\sim 10^6 \,M_\odot$~\cite{Lu:2025qbp}.
We note that along a similar vein, Ref.~\cite{Dekker:2024nkb} used the impact of modified small scale structure on the central density of dwarf galaxies to constrain the matter power spectrum, although their study assumed a power-law for the enhancement.

To summarize, in the enhanced power scenarios, these subhalos are more likely to be detected in stellar streams both due to their increased abundance \textit{and} due to the increased density of the subhalos raising the velocity kick from the impact.
A future survey such as the Via Project would then be sensitive to subhalos at least as small as $10^6 \,M_\odot$~\cite{2026arXiv260618332T}.
We leave a detailed study of the effect on Milky Way substructure and observability in stellar streams as the subject of future work.

\section{Conclusion}
\label{sec:conclude}

The baryonic consequences of enhanced density fluctuations remains relatively underexplored.
Since the halos resulting from large density fluctuations form early, they are denser and more compact, raising their virial temperature relative to halos in $\Lambda$CDM at the same mass.
This means that the halo mass threshold for forming astrophysical objects can be lower in these enhanced overdensities, and the high ambient dark matter density also leads to rich stellar dynamics.

In this work, we have explored the formation of stellar systems from enhanced density fluctuations.
We showed that the higher density of the resulting early halos raises their virial temperature, and therefore allows halos to form stars below the usual $\sim 10^8 \,M_\odot$ threshold.
If the effect of supernova feedback is also strong, then some of the halos that form stars will eject their stars and gas, leaving behind only a dark halo, with a density much higher than expected in $\Lambda {\rm CDM}$.
In addition, the dynamical friction timescales for stars in these dense halos can be short enough that stars can come to dominate the gravitational potential, even for relatively modest density fluctuations at wavenumbers above $k \sim 200$~Mpc$^{-1}$; hence, the stars in such halos will appear as old star clusters as opposed to old galaxies.
We also discuss various astrophysical uncertainties associated with these calculations, which alter the amplitudes of the different thresholds but not their scalings with wavenumber or halo mass.

For the fiducial enhanced power spectra considered in Ref.~\cite{Qin:2025ymc}, which produced enough NQPBHs to explain the number density of little red dots, we estimate the initial abundance of clusters formed through this dynamical friction mechanism.
We found that the initial number of clusters captured by the Milky Way is highly sensitive to how broad the enhancement in power is, with the narrow-peak enhancement resulting in 50 clusters within the Milky Way and the step-function case resulting in $4.1 \times 10^5$.
Given that there are many processes (e.g. evaporation and tidal shocks) that can efficiently disrupt globular clusters, it is possible that the large initial number of clusters in the latter case can be pared down to the $\sim 150$ globular clusters observed in the Milky Way, but we leave a detailed study of these destruction rates to future study.
We also briefly discussed the altered star formation history and possible contributions to the global ionization history. 

Moreover, as these early dense halos are subsumed within larger halos, the resulting dark matter substructure is altered from $\Lambda$CDM expectations.
Intermediate-mass subhalos ($10^7 - 10^9 \,M_\odot$) tend to be more compact and more abundant than in a standard cosmology, although the exact mass range where the enhancement appears depends on the shape of the matter power spectrum.
Both of these effects will enhance the detectability of small dark matter subhalos in upcoming stellar stream surveys.

The implications for baryonic structures forming in early and dense halos have not been comprehensively studied, but may provide multiple useful handles on yet-unmeasured scales in the matter power spectrum. 
The observables discussed in this work provide a few new directions with which to illuminate small-scale structure and motivate further study of the astrophysical consequences of enhanced structure formation.

\section*{Acknowledgments}
We thank Duncan Adams, Gus Beane, Charlie Conroy, Hitesh Das, David Hogg, Lachlan Lancaster, Adrian Price-Whelan, Katelin Schutz, Oren Slone, and Nadia Zakamska for useful discussions regarding this work.
W.Q. is supported by the Simons Society of Fellows through Grant No. SFI-MPS-SFJ-00006250. 
N.W. is supported by the National Science Foundation grant PHY-2210498, the Simons Foundation, and the US-Israel Binational Science Foundation (BSF) under grant 2022287.

\appendix

\section{Clustering time for different density profiles}
\label{app:dens_prof}

In Section~\ref{sec:hard_way}, we calculated the condition for stars to cluster densely within the lifetime of a halo by assuming the dark matter density profile was an isothermal sphere.
Here, we show that the condition derived there holds for other choices of density profiles.

\begin{figure}
    \centering
    \includegraphics[width=\columnwidth]{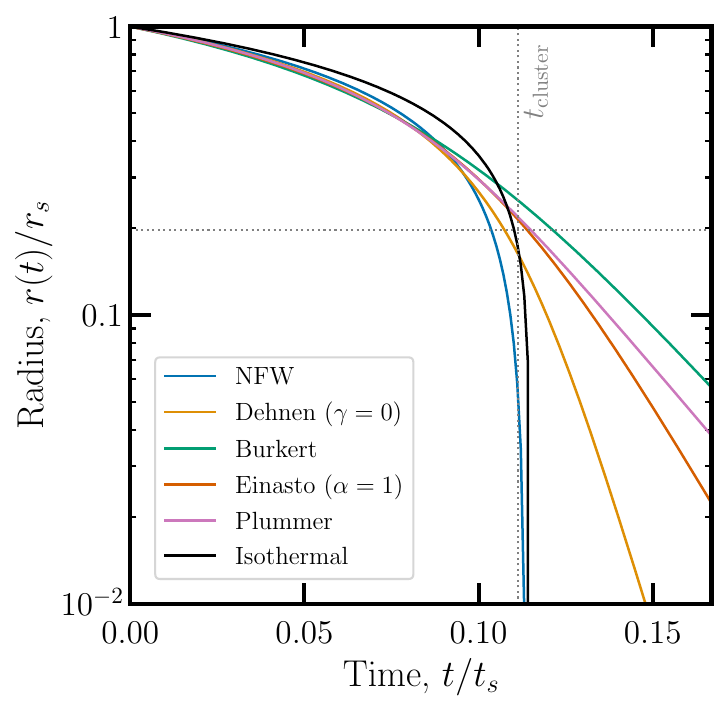}
    \caption{
        The trajectory of a particle under dynamical friction for different density profiles.
        The radius $r$ is in units of $r_s$, and $t$ is in units of $t_s = r_s^2 \sigma / (GM)$.
    }
    \label{fig:r_evolution}
\end{figure}
We numerically solve Eq.~\eqref{eq:r_evol} assuming the following different density profiles (in addition to the NFW profile, defined in Eq.~\eqref{eq:NFW}):
\begin{gather}
    \rho_{\rm Dehnen}(r) = \frac{(3-\gamma)\rho_s}{\left(\frac{r}{r_s}\right)^\gamma \left(1 + \frac{r}{r_s}\right)^{4-\gamma}} , \\
    \rho_{\rm Burkert}(r) = \frac{\rho_s}{\left(1 + \frac{r}{r_s}\right)\left(1 + \left(\frac{r}{r_s}\right)^2\right)} , \\
    \rho_{\rm Einasto}(r) = \rho_s \exp\left[-\frac{2}{\alpha}\left(\left(\frac{r}{r_s}\right)^\alpha - 1\right)\right] , \\
    \rho_{\rm Plummer}(r) = \frac{\rho_s}{\left(1 + \left(\frac{r}{r_s}\right)^2\right)^{5/2}} \\
    \rho_{\rm iso}(r) = \rho_s \frac{r_s^2}{r^2} .
\end{gather}
We begin integration at $r=r_s$ for each profile and choose $\rho_s$ such that $\rho(r_s)$ is the same across every profile.
For the Dehnen profile, we take $\gamma=0$ (as was done in Ref.~\cite{Graham:2023unf}), and for the Einasto profile, we use $\alpha=1$ to create a cored profile.

Fig.~\ref{fig:r_evolution} shows the result of solving Eq.~\eqref{eq:r_evol} for several different density profiles.
Time is shown on the $x$-axis and is normalized by $t_s = r_s^2 \sigma / (GM_*)$, where we reiterate that $M_*$ is the mass of a single star---in these units, the clustering time given in Eq.~\eqref{eqn:t_cluster} is $t_\mathrm{cluster} \sim 0.11 t_s$.
For a star beginning at a typical radius, we can see that the time it takes for a star to fall in to $r/r_0 \sim 0.2$ is quite similar across density profiles, with values falling between $t \in [0.10, 0.12] \times t_s$, and hence very near to $t_\mathrm{cluster}$.
It is therefore justified to use $t_\mathrm{cluster}$ as the clustering timescale no matter the density profile.

Intuitively, the similarity in clustering time across profiles is due to the fact that Eq.~\eqref{eq:r_evol} more directly on the \textit{mass density at $r$}, as opposed to the mass enclosed.
Since we solve this equation to determine when the radius changes by an $\mathcal{O} (1)$ factor, across which the density profiles also change only by an $\mathcal{O} (1)$ factor, then the result across different density profiles is similar.
If the central question was instead how long does it take a star to fall \textit{all the way to the center} to the halo, then the choice of profile becomes much more important due to the varying inner scalings of the density profiles, as can be seen from Fig.~\ref{fig:r_evolution} at small values of $r(t)$.

\bibliography{references}
\end{document}